\documentclass[preprint, 10pt]{aastex}       % For the LCD systems
\usepackage{xcolor}
\newcommand{\dern}[3][\;\;]{\ensuremath{ \frac{d^{#3}{#1}}{d{#2}^{#3}} }}

\newcommand{\bvec}[1]{{\mbox{{\boldmath$#1$}}}}		% bold in mathmode
\newcommand{\unitv}[1]{\bvec{\hat{#1}}}			% bold unit vector
\newcommand{\jfm}{JFM}		%Journal of Fluid Mechanics
\newcommand{\natas}{NatAs}	%Nature Astronomy
\newcommand{\ptrslA}{RSPTA}	%Philosophical Transactions of the Royal Society of London Series A
\newcommand{\rsptA}{RSPTA}	%Philosophical Transactions of the Royal Society of London Series A
\newcommand{\pnas}{PNAS}	%Proceedings of the National Academy of Science
\newcommand{\sci}{Sci}		%Science
\newcommand{\scia}{SciA}	%Science Advances
\newcommand{\eqnref}[1]{(\ref{#1})}
\begin{document}

\title{Overstable Convective Modes in a Polytropic Stellar Atmosphere}
\shorttitle{Overstable Convective Modes in a Polytropic Stellar Atmosphere}

\author{Bradley W. Hindman}
\affil{JILA, University of Colorado, Boulder, CO~80309-0440, USA}
\affil{Department of Applied Mathematics, University of Colorado, Boulder, CO~80309-0526, USA}
\email{hindman@solarz.colorado.edu}

\author{Rekha Jain}
\affil{School of Mathematics \& Statistics, University of Sheffield, Sheffield S3 7RH, UK}

%%%%%%%%%%%%%%%%%%%%%%%%%%%%%%%%%%%%%%%%%%%%%%%%%%%%%%%%%%%%%%%%%%%%%%%%%%%%%%%%%
%                                                                              	%
%               Abstract                                                       	%
%										%
%%%%%%%%%%%%%%%%%%%%%%%%%%%%%%%%%%%%%%%%%%%%%%%%%%%%%%%%%%%%%%%%%%%%%%%%%%%%%%%%%

\begin{abstract}

Within the convection zone of a rotating star, the presence of the Coriolis force
stabilizes long-wavelength convective modes. These modes, that would have been
unstable if the star lacked rotation, are called overstable convective modes or
thermal Rossby waves. We demonstrate that the Sun's rotation rate is sufficiently
rapid that the lower half of it's convection zone could possess overstable modes.
Further, we present an analytic solution for atmospheric waves that reside within
a polytropic stratification. We explore in detail the properties of the overstable
and unstable wave modes that exist when the polytrope is weakly unstable to convective
overturning. Finally, we discuss how the thermal Rossby waves that reside within
the convection zone of a star might couple with the prograde branch of the $g$ modes
that are trapped within the star's radiative zone. We suggest that such coupling
might enhance the photospheric visibility of a subset of the Sun's $g$ modes.

\end{abstract}


%%%%%%%%%%%%%%%%%%%%%%%%%%%%%%%%%%%%%%%%%%%%%%%%%%%%%%%%%%%%%%%%%%%%%%%%%%%%%%%%%
%										%
%       1.      Introduction							%
%										%
%%%%%%%%%%%%%%%%%%%%%%%%%%%%%%%%%%%%%%%%%%%%%%%%%%%%%%%%%%%%%%%%%%%%%%%%%%%%%%%%%

\section{Introduction}
\label{sec:Introduction}

The Coriolis force acts to inhibit convection, in both obvious and subtle ways. In
addition to suppressing the efficiency of turbulent heat transport \citep[e.g.,][]
{Julien:1996, Brandenburg:2009}, reducing the dominant spatial scale of the convective
motions \citep[e.g.,][]{Featherstone:2016b, Vasil:2021}, and inducing anisotropy
and latitudinal variation in the convective heat flux \citep[e.g.,][]{Cowling:1951,
Tayler:1973}, rotation has long been known to delay the onset of the convective
instability, increasing the thermodynamic gradients required for spontaneous convective
overturning. In a Boussinesq system this appears as an increase in the critical
Rayleigh number that marks the onset of the instability \citep[see][]{Chandrasekhar:1961}.
In a gravitationally stratified fluid, the local Ledoux criterion \citep{Ledoux:1947}
is modified by rotation such that a steeper specific entropy or compositional gradient
is required for instability. For example, \cite{Cowling:1951} argued that non-axisymmetric
motions are locally stable if the square of the buoyancy frequency $N^2$ exceeds a
rotationally dependent threshold (which is negative),

\begin{equation} \label{eqn:Cowlings_Criterion}
	N^2 > -4 \Omega^2 \frac{k_\Omega^2}{k_h^2} \; .
\end{equation}

\noindent In Cowling's criteria, $\Omega$ is the star's rotation rate and $k_\Omega$
and $k_h$ are two components of the wavenumber vector of the convective waveform.
The component parallel to the rotation axis is $k_\Omega$ and the horizontal component,
perpendicular to gravity, is $k_h$. This criterion makes it clear that certain modes
of convection (i.e, certain wavenumbers) can be stabilized by rotation even if the
atmosphere is globally unstable by the Ledoux criteria, namely $N^2 < 0$.

Cowling's formula further suggests that rotation only influences the stability of
those convective motions that have variation parallel to the rotation axis, i.e.,
$k_\Omega \neq 0$. Without such variation, Cowling's criterion reduces to that of
Ledoux, $N^2 > 0$. We now know that Cowling's analysis was incomplete. An entire
class of gravito-inertial waves was excluded from the dispersion relation that was
used to derive the stability criterion. These waves, called thermal Rossby waves in
geophysics and planetary science and referred to as overstable convective modes or
low-frequency prograde waves in astrophysics \citep[see][]{Ando:1989, Unno:1989},
are essentially convective modes that have been stabilized and made oscillatory by
rotation. Stable forms of these waves exist that disobey Cowling's criterion. The
existence of such waves requires either curvature of the boundaries to provide a
topological $\beta$-effect \cite[e.g.,][]{Roberts:1968, Busse:1970, Busse:1986} or radial
stratification of the density producing a compressional $\beta$-effect \citep[e.g.,][]
{Hide:1966, Glatzmaier:1981,Unno:1989, Hindman:2022}. Both cases enable a
prograde-propagating vorticity wave because a spinning convective column that is
pushed toward the axis of rotation must spin faster either from vortex stretching
induced by the shape of the boundaries or by vortex narrowing caused by density
stratification.

Such thermal Rossby waves are absent from Cowling's calculation because his analysis
was a purely local one that ignored the presence of physical boundaries and did not
consider spatial variation in the atmosphere beyond the term involving the buoyancy
frequency. \cite{Ando:1989} expanded on Cowling's derivation by deriving a local
dispersion relation that includes density  stratification. From this dispersion
relation one can derive the following local stability criterion for overstable
convective modes,

\begin{equation} \label{eqn:Andos_Criterion}
	N^2 > -4 \Omega^2 \frac{k_\Omega^2}{k_h^2} - \frac{\Omega^2\sin^2\theta}{{\cal H}^2}
		\frac{k_\phi^2}{k_h^2 k^2} \; ,
\end{equation}

\noindent where $k_\phi$ is the azimuthal component of the wavenumber, $k$ is the
total wavenumber, $\theta$ is the colatitude, and $\cal H$ is a length scale that
depends on the density scale height $H$, the buoyancy frequency $N$, and the
gravitational acceleration $g$,

\begin{equation}
	\frac{1}{\cal H} \equiv \frac{1}{H} - \frac{2N^2}{g} \; .
\end{equation}

\noindent In a stellar convection zone, the buoyancy frequency is small and this
scale length is nearly equal to the density scale height ${\cal H} \approx H$. If
the wavelength is much shorter that the density scale height ($kH \gg 1$), the
criterion described by Equation~\eqnref{eqn:Andos_Criterion} reduces to that of
Cowling, Equation~\eqnref{eqn:Cowlings_Criterion}, unless the wavenumber is purely
perpendicular to the rotation axis ($k_\Omega=0$). For waves with such an alignment,
Ando's correction to Cowling's criterion becomes the dominant term. If one follows
\cite{Hindman:2022} and includes the effect of the acoustic cut-off frequency,
$\omega_c$, the stability condition for thermal Rossby waves near the equator
($\sin\theta\approx 1$) becomes

\begin{equation} \label{eqn:stability}
	N^2 > - \frac{\Omega^2}{(k^2 + k_c^2){\cal H}^2} \; ,
\end{equation}

\noindent where $k_c$ is a cut-off wavenumber that depends solely on the density
scale height and its radial derivative,

\begin{equation}
	k_c^2 \equiv \frac{\omega_c^2}{c^2} = \frac{1}{4H^2} \left(1-2\frac{dH}{dr}\right) \; ,
\end{equation}

\noindent where $c$ is the sound speed. The cut-off wavenumber encapsulates the
reflection that can occur when a wave travels vertically into a region where its
vertical wavelength becomes long compared to the density height. As such, the inclusion
of the cut-off wavenumber is particularly important for the long wavelength waves
for which $kH \ll 1$. In this limit, the stability criteria becomes exceedingly
simple. For example, deep within a stellar convection zone where ${\cal H} \approx H$
and $ k_c^2 \approx 1/4H^2$, we find stability for $N^2> -4\Omega^2$. 

\begin{figure*}
	\epsscale{0.5}
	\plotone{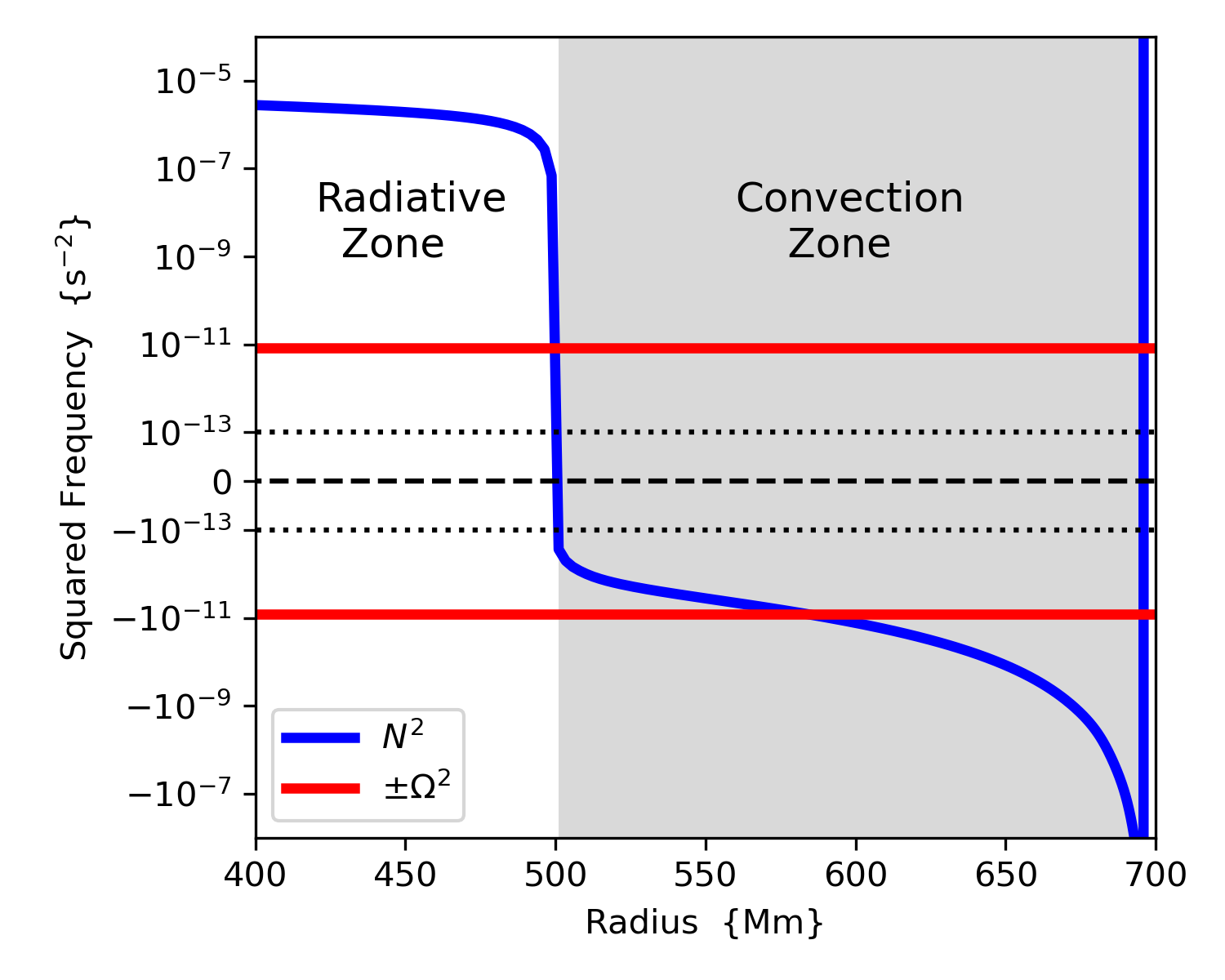}

	\caption{Square of the buoyancy frequency (blue curve) within the the Sun's
convection zone as achieved in Model S \citep{Christensen-Dalsgaard:1996}, a standard
model of the Sun's internal structure. The ordinate axis is scaled logarithmically
for both positive and negative values that have magnitudes greater than $10^{-13}$ s$^{-2}$.
In the region between the two black dotted lines, $\left|N^2\right| < 10^{-13} ~{\rm s}^{-2}$,
the scaling is linear. The two red horizontal lines indicate plus and minus the square
of the Sun's Carrington rotation rate, $\pm\Omega_\odot^2$, where
$\Omega_\odot = 2.87 \times 10^{-6} ~{\rm s}^{-1}$. In the lower half of the convection zone,
the rotation is large enough to stabilize long-wavelength convective modes
($\Omega_\odot > \left|N\right|$); thus, convectively unstable modes are converted into
overstable gravito-inertial waves.
	\label{fig:modelS}}
\end{figure*}

How likely are stable thermal Rossby waves to exist in a star like the Sun? From an
examination of Equation~\eqnref{eqn:stability} it becomes clear that stable waves can
exist in a nominally unstable stratification ($N^2 < 0$) as long as the rotation rate
is of the same order as (or larger than) the modulus of the buoyancy frequency.
Figure~\ref{fig:modelS} illustrates the square of the buoyancy frequency $N^2$ as
a function of radius within model S, a standard model of the Sun's interior structure
\citep{Christensen-Dalsgaard:1996}. The square of the Sun's Carrington rotation rate,
$\Omega_\odot^2$, is indicated by the horizontal red lines. Within the Sun's convection
zone (shaded gray), the square of the buoyancy frequency is negative and thus the
atmosphere is globally unstable to convection. However, within the lower half of the
Sun's convection zone, the rotation rate becomes larger than the modulus of the buoyancy
frequency and stable thermal Rossby waves become possible for long wavelengths. \cite{Hindman:2022}
estimated that waves with an azimuthal order $m=k_\phi R_\sun$ less than 30 might be
stabilized by the rotation.

Since thermal Rossby waves are low frequency they will have frequencies commensurate
with a subset of the $g$ modes that reside in the Sun's radiative interior. Thus,
there exists the intriguing possibility that the thermal Rossby waves will be
coupled with the prograde branch of the $g$ modes. In $\beta$ Cephei stars, the
coupling of thermal Rossby waves that reside within the convective core with the
$g$ modes of the overlying stable envelope has been proposed as a mechanism for
the excitation of pulsations \citep{Osaki:1974, Lee:1986, Lee:1987}. But, such
coupling has not been explored previously in a solar context.

The nonlinear waveforms, often called banana cells, that result when thermal 
Rossby waves are unstable have been studied extensively in the solar and stellar
context through numerical simulations \citep[e.g.,][]{Miesch:2000, Hotta:2015,
Nelson:2018, Hindman:2020a,Hindman:2020b}. While a few studies of waves in
atmospheres with adiabatic stratification have been performed in the past
\citep{Glatzmaier:1981, Bekki:2022, Hindman:2022}, far less attention has been
paid to the stable form of these waves when the atmosphere is convectively unstable
($N^2 < 0$). This  will be our goal here, to derive the eigenfrequencies and
eigenfunctions for the thermal Rossby waves that exist in a weakly unstable
stratification. In particular, we explore the analytic solution presented by
\cite{Hindman:2022} for thermal Rossby waves within a polytropic atmosphere. That
solution is valid irrespective of whether the polytropic stratification is stable,
unstable, or neutrally stable. Here we examine the potentially complex eigenfrequencies
and eigenfunctions that apply for globally unstable stratifications.

Section 2 of this paper provides the governing equation for the thermal Rossby waves
within a stratified atmosphere. Section 3 describes the polytropic atmosphere and
presents the analytic solutions. Section 4 explores how the thermal Rossby waves
that reside within the solar convection zone might couple with the $g$ modes that
are trapped within the radiative interior. Finally, in Section 5 we discuss the
implications of our results on mode stability and the visibility of $g$ modes.

%%%%%%%%%%%%%%%%%%%%%%%%%%%%%%%%%%%%%%%%%%%%%%%%%%%%%%%%%%%%%%%%%%%%%%%%%%%%%%%%%
%										%
%       2.    Atmospheric Waves in a Rotating Star				%
%										%
%%%%%%%%%%%%%%%%%%%%%%%%%%%%%%%%%%%%%%%%%%%%%%%%%%%%%%%%%%%%%%%%%%%%%%%%%%%%%%%%%

\section{Atmospheric Waves in a Rotating Star}
\label{sec:Atmospheric_Waves}

We seek solutions for sectoral modes that propagate longitudinally and potentially
radially. We prohibit propagation in the latitudinal direction and assume that any
atmospheric variation in that direction can be ignored. Such a 2D approximation is
consistent with waves that are rotationally dominated and satisfy the Taylor-Proudman
constraint. Further, we ignore the curvature of the star's isopycnals and
assume constant gravity within a local plane-parallel model. Since we are interested in
sectoral modes, we place the local Cartesian coordinate system at the star's equator.
We orient the axes of this coordinate system such that $\unitv{x}$ points in the
longitudinal direction, $\unitv{z}$ points radially and antiparallel to the gravitational
acceleration $\bvec{g} = - g\unitv{z}$, and $\unitv{y}$ points in the invariant
latitudinal direction, parallel to the star's rotation vector,
$\bvec{\Omega} = \Omega \unitv{y}$. Under such conditions, \cite{Hindman:2022}
have shown that linear atmospheric waves in the rotating reference frame satisfy
the following governing equation, 

\begin{equation}
	\label{eqn:governing_equation}
	\dern[\Psi]{z}{2} + k_z^2(z) \Psi(z) = 0 \; , 
\end{equation}

\noindent where $k_z$ is a height-dependent vertical wavenumber,

\begin{equation}
	k_z^2(z) \equiv \frac{\omega^2 - \omega_c^2 - 4\Omega^2}{c^2} 
			- k_x^2 \left(1 - \frac{N^2}{\omega^2}\right)
			+ \frac{2\Omega k_x}{\omega}\left(\frac{1}{H} - \frac{2N^2}{g}\right) \; .
\end{equation}

This local vertical wavenumber depends on the atmospheric profiles of the sound
speed $c$, buoyancy frequency $N$, acoustic cutoff frequency $\omega_c$, and density
scale height $H$. The variable $\Psi(z)$ expresses the radial behavior of the wave's
Lagrangian pressure fluctuation $\delta P$, scaled by the square root of the atmosphere's
mass density $\rho_0$,

\begin{equation}
	\delta P(x,z,t) = \rho_0^{1/2} \, \Psi(z) \, e^{i\left(k_x x - \omega t\right)} \; .
\end{equation}

\noindent We seek plane-wave solutions in the longitudinal direction where $\omega$
is the temporal frequency and $k_x$ is the zonal wavenumber. We have adopted the
sign convention that for a positive wavenumber, $k_x > 0$, the waves propagate
in the prograde direction if the frequency is positive, $\omega > 0$, and in the
retrograde direction for negative frequencies, $\omega < 0$. For comparison to
waves in spherical geometry, the zonal wavenumber is related to the azimuthal order
of the concomitant spherical harmonic $m = k_x R$, where $R$ is the stellar
radius. Equation~\eqnref{eqn:governing_equation} describes the propagation of acoustic
waves and gravito-inertial waves within an atmosphere with a general stratification.
In the next section we specialize to a polytropic stratification which is particularly
relevant to stellar convection zones.

%%%%%%%%%%%%%%%%%%%%%%%%%%%%%%%%%%%%%%%%%%%%%%%%%%%%%%%%%%%%%%%%%%%%%%%%%%%%%%%%%
%										%
%       3.    Waves Modes for a Polytropic Stratification			%
%										%
%%%%%%%%%%%%%%%%%%%%%%%%%%%%%%%%%%%%%%%%%%%%%%%%%%%%%%%%%%%%%%%%%%%%%%%%%%%%%%%%%
\section{Polytropic Stratification}
\label{sec:Polytropic_Stratification}

A polytrope is an atmosphere for which the atmospheric pressure $P_0$ is related
to the mass density $\rho_0$ through a power-law relation, where the exponent is
usually expressed in terms of a polytropic index $\alpha$,

\begin{equation}
	P_0 \sim \rho_0^{(\alpha+1)/\alpha} \; .
\end{equation}

\noindent In such an atmosphere all of the thermodynamic profiles become power-law
functions of the height coordinate, $z$. Hence, the pressure, density, and temperature
all vanish at the height $z=0$ and the atmosphere exists only within the half-space
$z<0$. For a polytropic stratification, the governing equation can be rewritten in
the form of Whittaker's Equation \citep[for details see][]{Hindman:2022},

\begin{equation} \label{eqn:Whittaker}
	\dern[\Psi]{\zeta}{2} + 
		\left[ \frac{\kappa}{\zeta} - \frac{1}{4} - \frac{\nu(\nu-1)}{\zeta^2} \right] \Psi = 0 \; ,
\end{equation}

\noindent where $\zeta = -2 k_x z$ is a dimensionless depth, $\nu \equiv \left(\alpha + 2\right)/2$
depends on the stratification, and $\kappa$ is a constant eigenvalue,

\begin{eqnarray}
	\kappa &\equiv& \textcolor{red}{\left(\frac{\alpha+1}{2\gamma}\right) \frac{\omega^2-4\Omega^2}{gk_x}}
		+ \textcolor{blue}{\frac{\alpha-\hat{\alpha}}{2\gamma\hat{\alpha}}\frac{gk_x}{\omega^2}}
		+ \textcolor{orange}{\left[\alpha - \frac{2\left(\alpha-\hat{\alpha}\right)}{\gamma\hat{\alpha}}\right] \left(\frac{\Omega}{\omega}\right) } \; .
	\label{eqn:definition_kappa}
\end{eqnarray}

\noindent In the expression for $\kappa$ above, $\gamma$ is the fluid's adiabatic
exponent and $\hat{\alpha}$ is the value of the polytropic index that corresponds
to an atmosphere that is neutrally stable to convective overturning,
$\hat{\alpha} \equiv \left(\gamma-1\right)^{-1}$. If $\alpha > \hat{\alpha}$ the
atmosphere is stably stratified and if $\alpha<\hat{\alpha}$ the atmosphere is unstable
to convective motions. The constant $\kappa$ is the radial eigenvalue of the ODE.
Since $\kappa$ depends on the frequency and zonal wavenumber, once any given eigenvalue
is obtained by applying boundary conditions, Equation~\eqnref{eqn:definition_kappa}
serves as a global dispersion relation for that radial mode. The term that is
colored red in Equation~\eqnref{eqn:definition_kappa} engenders acoustic oscillations
and arises from the compressibility of the fluid with a weak correction by the Coriolis
force. The blue term is due to buoyancy and is responsible for internal gravity waves.
Finally, the orange term results from the Coriolis force and produces inertial waves.

Whittaker's Equation \citep{Abramowitz:1964} has two solutions that can be expressed
in terms of Kummer's confluent hypergeometric functions of the first and second kind,
$M$ and $U$. The general solution for the Lagrangian pressure fluctuation is therefore
a linear combination of these two solutions,

\begin{equation}
	\delta P(z) = \rho_0^{1/2} \, \Psi(z) = z^{\alpha+1} \, e^{k_x z}
		\left[ C_M M\left(\nu-\kappa, 2\nu, -2k_x z\right) +
		       C_U U\left(\nu-\kappa, 2\nu, -2k_x z\right)\right] \; ,
\end{equation}

\noindent with arbitrary constants $C_M$ and $C_U$ whose ratio is determined by the
boundary conditions. We will examine two different sets of boundary condition,
one that is appropriate for a semi-infinite domain, $z\in(-\infty,0\,]$ and
the other for a finite domain with a depth of $D$, $z\in[-D,0\,]$.

%-------------------------------------------------------------------------------%
%	3.1	Eigenvalues for a Semi-infinite Domain

\subsection{Eigenvalues for a Semi-infinite Domain}
\label{subsec:semi-infinite_domain}

For illustrative purposes, we will consider a semi-infinite domain where we impose
regularity conditions at the two singular points of Whittaker's equation (i.e., at
$z=0$ and $z\to-\infty$). Of course, a real star does not possess an infinite domain.
But, as long as the turning points that correspond to the edges of the wave cavity
are far from the physical boundaries, the solution is  insensitive to whether we
apply the boundary conditions at a finite depth or not. From Whittaker's
Equation~\eqnref{eqn:Whittaker} we can deduce that deep in the polytropic atmosphere
($z \to -\infty$ or $\zeta \to \infty$) the solutions are evanescent and decay
exponentially with the leading behavior, $\Psi \sim \exp\left(k_x z\right)$. Further,
by setting the coefficient in square brackets in Equation~\eqnref{eqn:Whittaker}
to zero, we find that the wave cavity has a lower boundary (or turning point) that
predominantly scales inversely with the zonal wavenumber,

\begin{equation}
	z_{\rm turn} = -\frac{\zeta}{2k_x} =
	    k_x^{-1}\left[-\kappa + \left(\kappa^2 - \nu^2+\nu\right)^{1/2}\right] \; .
\end{equation}

\noindent Hence, for short zonal wavelengths, the wave cavity is confined within
the upper portion of the polytrope and the eigenvalues are insensitive to a boundary 
condition that is applied at a deep but finite depth.

When the previously discussed conditions of regularity are imposed, the eigenvalue
$\kappa$ is restricted to a discrete spectrum of values labelled by their radial
order $n$,

\begin{equation}
	\kappa = \kappa_n = n + \frac{\alpha + 2}{2} \; ,
\end{equation}

\noindent which can be any non-negative integer, $n \in 0,1,2,3, \cdots$. Note, the
eigenvalue is insensitive to the zonal wavenumber $k_x$ because a semi-infinite 
polytropic atmosphere is self-similar and lacks an imposed scale length
\citep[see][]{Hindman:2022}. In addition to the simple analytic expression for the
eigenvalue, the eigenfunctions also become more tractable. Regularity at both singular
points collapses both Kummer functions into Associated Laguerre Polynomials
\citep{Abramowitz:1964}. The solution for the eigenfunction of the $n$th radial
order is therefore given by

\begin{equation}
	\label{eqn:infinite_eigenfunctions}
	\delta P_n(z,x,t) = C_n \, z^{\alpha + 1} \, e^{k_x z} \, L_n^{(\alpha+1)}\left(-2k_xz\right) \, e^{i(k_x x - \omega_n t)} \; ,
\end{equation}

\noindent where $C_n$ is an arbitrary constant and $L_n^{(a)}$ is the $n$th-order
Associated Laguerre Polynomial. In sections~\ref{subsec:slow_rotation} and
\ref{subsec:eigenfunctions} we will discuss the resulting eigenfrequencies
and eigenfunctions in conjunction with those appropriate for the finite domain
that we will discuss in the next subsection.

%-------------------------------------------------------------------------------%
%	3.2	Eigenvalues for a Finite Domain

\subsection{Eigenvalues for a Finite Domain}
\label{subsec:finite_domain}

In this subsection we model the convection zone of a low-mass star as a polytropic
layer of finite depth $D$, spanning $z \in [-D,0\,]$. For boundary conditions, we
impose regularity at the origin (a singular point of Whittaker's Equation) and vanishing
of the Lagrangian  pressure fluctuation at the bottom of the layer---which for low
frequency is consistent with a condition of impenetrability \citep[see][]{Hindman:2022}.
Since, the $U$ hypergeometric function is singular at the origin, the eigenfunction
only depends on the $M$ confluent hypergeometric function,

\begin{equation}
	\delta P_n(z,x,t) = C_n \, z^{\alpha + 1} \, e^{k_x z} \,
		M\left(\nu - \kappa_n, 2\nu, -2k_x z\right) \,
		e^{i(k_x x - \omega t)} \; ,
\end{equation}

\noindent and the eigenvalue $\kappa_n$ is determined by a transcendental dispersion
relation that enforces the boundary condition at the bottom of the layer,

\begin{equation}
	M\left(\nu - \kappa_n, 2\nu, 2k_x D\right) = 0 \; .
\end{equation}

\begin{figure*}
	\epsscale{0.5}
	\plotone{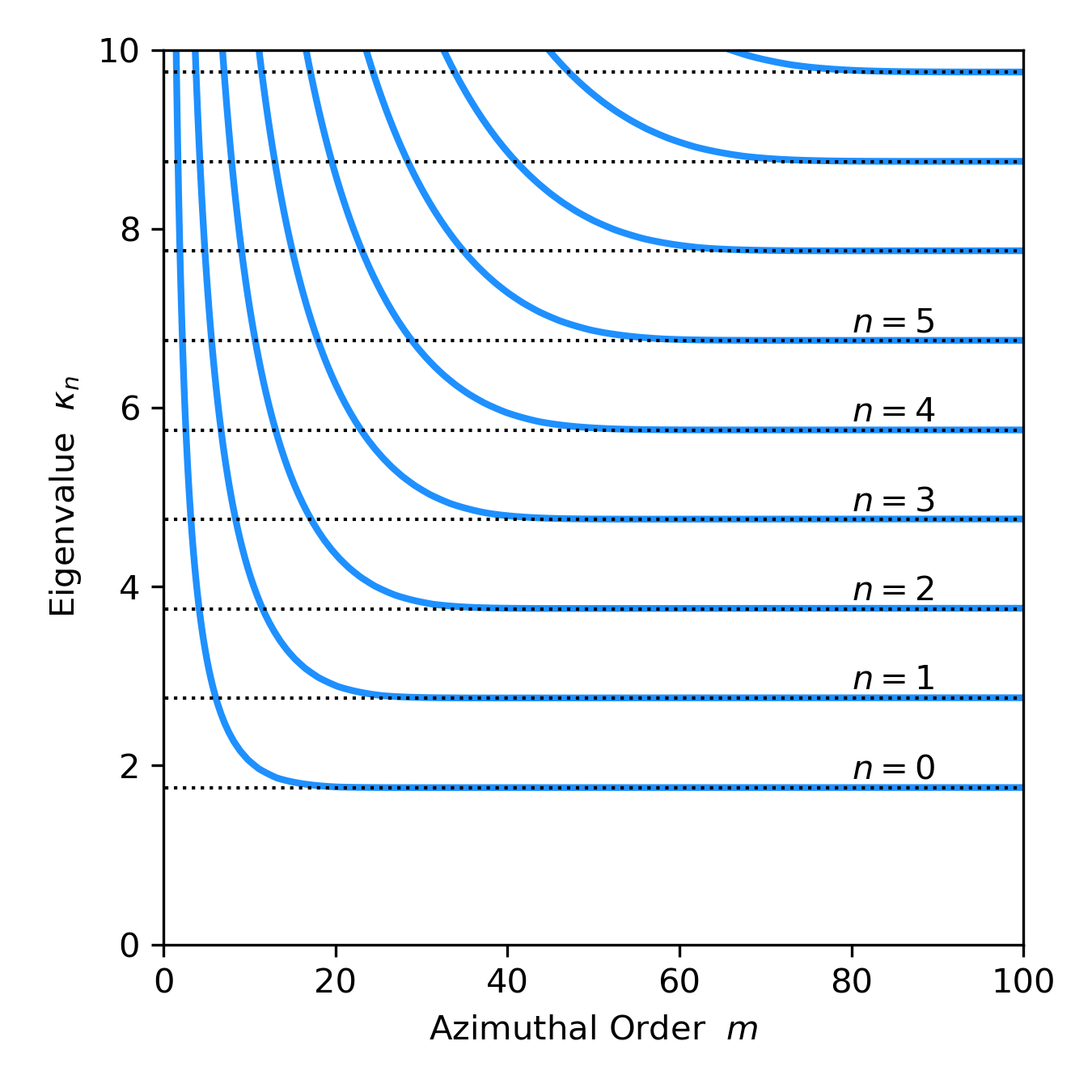}

	\caption{The eigenvalue $\kappa$ as a function of the azimuthal order $m$.
The horizontal dotted lines indicate the eigenvalue for the semi-infinite domain
discussed in section~\ref{subsec:semi-infinite_domain}. Each line corresponds
to a different radial order $n$, $\kappa_n = n + 1 + \alpha/2$. The solid blue
curves show the eigenvalue for the finite domain presented in
section~\ref{subsec:finite_domain}. The two domains have eigenvalues that converge
to a common value for large zonal wavenumber.
	\label{fig:eigenvalues}}
\end{figure*}

\noindent We have numerically solved for the roots of this equation for a convection
zone depth of $D=200$ Mm and a stellar radius equal to that of the Sun, $R = 696$ Mm.
Figure~\ref{fig:eigenvalues} illustrates the resulting eigenvalues $\kappa_n$ as
a function of azimuthal order $m = k_x R$ for modes with a radial order $n$ less
than 9. The solid blue curves show $\kappa_n$ for the finite domain, whereas the
black dotted lines illustrate the eigenvalue for the semi-infinite domain,
$\kappa_n = n + \alpha/2 + 1$. The eigenvalues for the finite domain begin with a large
value at low zonal wavenumber and decrease with increasing wavenumber, eventually
asymptoting to a constant value that corresponds to the eigenvalue that holds for
the semi-infinite domain. This asymptotic behavior occurs because the lower turning
point moves higher  in the atmosphere as the zonal wavenumber increases and once
it has passed into the domain (i.e., when $z_{\rm turn} > -D$), the boundary condition
at the bottom of the convection zone becomes increasing unimportant.

%-------------------------------------------------------------------------------%
%	3.3	Eigenfrequencies for Slow Rotation				%
%-------------------------------------------------------------------------------%
\subsection{Eigenfrequencies for Slow Rotation}
\label{subsec:slow_rotation}

Given a numerical value for the eigenvalue $\kappa$, Equation~\eqnref{eqn:definition_kappa}
can be solved to obtain the corresponding eigenfrequencies. In order to understand
the relative importance of each of the terms that appear in the definition of $\kappa$,
it is useful to define several nondimensional parameters.

\begin{eqnarray}
	\tilde{\omega} &\equiv& \frac{\omega}{\Omega} \; ,
\\
	A &\equiv& \frac{\left(\alpha+1\right)}{2\gamma} \; ,
\\
	\label{eqn:S_def}
	S &\equiv& \frac{\alpha-\hat{\alpha}}{2\gamma\hat{\alpha}}\frac{g}{\Omega^2 R} \; ,
\\
	\epsilon &\equiv& \frac{\Omega^2 R}{g} \; .
\end{eqnarray}

\noindent The nondimensional wave frequency $\tilde{\omega}$ is based on the star's
rotational period and $A$ is a coefficient for the term that leads to acoustic waves. 
The quantity $S$ acts as a convective stability criterion for the stratification. When
$S>0$ the atmosphere is stable to convective overturning and when $S<0$ the atmosphere
is unstably stratified. The magnitude of $S$ indicates the importance of buoyancy to
rotation. Since the buoyancy frequency, $N$, diverges at the origin of a polytrope
(where the density vanishes), there is always a region in the upper portion of the
polytrope where $\left|N\right| > \Omega$. Conversely, deep in the atmosphere the
buoyancy frequency vanishes as the depth increases and is therefore small compared to
the rotation rate, $\left|N\right| < \Omega$. The stability parameter $S$ determines
the depth of the transition between these two regimes, $z_{\rm tran} = -2SR$.
The parameter $\epsilon$ is a nondimensional measure of the rotational speed that
characterizes the centrifugal deformation of a star. For a star rotating much slower
than its break up rotation rate, $\epsilon$ is a small parameter. For example, for the
Sun $\epsilon = 2 \times 10^{-5}$ \citep{Gizon:2016}. The smallness of $\epsilon$ will
prove useful, enabling the expansion of the dispersion relations into distinct branches
for the acoustic waves and the gravito-inertial waves.

By using these nondimensional numbers, the definition of $\kappa$ can be rewritten
more succinctly

\begin{equation} \label{eqn:global_dispersion}
	\kappa_n = \textcolor{red}{\frac{A\epsilon}{m}\left(\tilde{\omega}^2-4\right)}
		+ \textcolor{blue}{\frac{mS}{\tilde{\omega}^2}}
		+ \textcolor{orange}{\frac{\alpha - 4 \epsilon S}{\tilde{\omega}}} \; .
\end{equation}

\noindent Once again, the red, blue, and orange terms are due to compressibility,
buoyancy, and the Coriolis force, respectively. This expression, when multiplied
by $\tilde{\omega}^2$, is a quartic polynomial in the frequency. Hence, for any given
value for the eigenvalue $\kappa_n$, there are four solutions for the frequency.
Furthermore, since the equation has the form of a depressed quartic and the coefficients
are all real-valued, two of the roots always have a real frequency and the other two
are either real or form a complex conjugate pair. The two roots that are always real
correspond to high-frequency acoustic waves and these arise from the red term in
Equation~\eqnref{eqn:global_dispersion}. The other two solutions (with real or complex 
frequencies) are low-frequency gravito-inertial waves and they arise jointly from the
blue and orange terms in Equation~\eqnref{eqn:global_dispersion}.

When $\epsilon$ is small (slow rotation), the high-frequency and low-frequency
branches can be easily separated by expanding the eigenfrequency in powers of the
small parameter $\epsilon$. The high-frequency solutions are found by expanding in
terms of $\epsilon^{1/2}$, 

\begin{equation}
	\tilde{\omega}_n = \epsilon^{-1/2} \sum_{j=0}^\infty a_j \epsilon^{j/2} \; .
\end{equation}

\noindent By solving the global dispersion relation~\eqnref{eqn:global_dispersion}
order by order, we find the first two terms in the expansion,

\begin{equation}
	\tilde{\omega}_n = \left(\frac{m\kappa_n}{A}\right)^{1/2} \epsilon^{-1/2}
		- \frac{\alpha}{2\kappa_n}
		+ {\cal O}\left(\epsilon^{1/2}\right) \; .
\end{equation}

\noindent In dimensional variables, this translates to

\begin{equation} \label{eqn:acoustic_frequencies}
	\omega_n = \pm \left(\frac{2\gamma}{\alpha + 1} \kappa_n g k_x\right)^{1/2} - \frac{\alpha\Omega}{2\kappa_n} + {\cal O}\left(\Omega \epsilon^{1/2}\right) \; .
\end{equation}

\noindent As expected, the eigenfrequencies for the acoustic waves are always real
and hence the acoustic modes are always stable.

The form of Equation~\eqnref{eqn:acoustic_frequencies} is agnostic of the boundary
conditions. The eigenfrequencies depend on the boundaries only through the eigenvalue
$\kappa$. If we employ the solution for $\kappa_n$ that applies for a semi-infinite
atmosphere,

\begin{equation}
	\kappa_n = n + \nu = n' + \alpha/2 \; ,
\end{equation}

\noindent with $n'=n+1$, we obtain the well-known eigenfrequencies for $p$ modes
in a polytropic atmosphere \citep{Lamb:1945} but with a rotational correction,

\begin{equation}
	\omega_n = \pm \left(\frac{\gamma\alpha}{\alpha + 1}\right)^{1/2}\left(\frac{2n'+\alpha}{\alpha} \, g k_x\right)^{1/2} - \frac{\alpha\Omega}{2n'+\alpha} + {\cal O}\left(\Omega \epsilon^{1/2}\right) \; .
\end{equation}

\noindent The mode with radial order $n$ produces the $p$ mode called $p_{n'}$,
i.e., $n=0$ corresponds to $p_1$, $n=1$ to $p_2$, and so on.

The low-frequency gravito-inertial waves are obtained with an expansion in powers
of $\epsilon$,

\begin{equation}
	\tilde{\omega} = \sum_{j=0}^\infty b_j \, \epsilon^j \; ,
\end{equation}

\noindent which to leading order produces

\begin{equation}
	\omega_n = \pm\frac{\alpha\Omega}{2\kappa_n} \pm \left(\frac{\alpha^2\Omega^2}{4\kappa_n^2}
		+ S \, \frac{gk_x}{\kappa_n}\right)^{1/2} + {\cal O}\left(\Omega \epsilon\right) \; .
\end{equation}

\noindent We immediately deduce that the eigenfrequencies are complex if the
atmosphere possesses a sufficiently strong unstable stratification,

\begin{equation}
	S < -\frac{\alpha^2\Omega^2}{4gk_x\kappa_n} = -\frac{\alpha^2\epsilon}{4 m \kappa_n} \; .
\end{equation}

\noindent The condition for stability is a global one that depends not only on the stratification, 
rotation rate, and wavelength, but also on the boundary conditions through the eigenvalue
$\kappa$. Even for an atmosphere with an unstable stratification ($S < 0$ or $\alpha < \hat{\alpha}$),
a mode can be stabilized by rotation if the mode has sufficiently low zonal wavenumber $k_x$
and low radial order $n$ (i.e., low $m$ and low $\kappa_n$). Equivalently, for a given radial
order $n$, stability parameter $S$, and rotation rate $\Omega$, the mode is stable for low
zonal wavenumbers and unstable for azimuthal orders exceeding a threshold,

\begin{equation}
	m > -\frac{\alpha^2 \epsilon}{4\kappa_n S} \; .
\end{equation}

\begin{figure*}
	\epsscale{1.0}
	\plotone{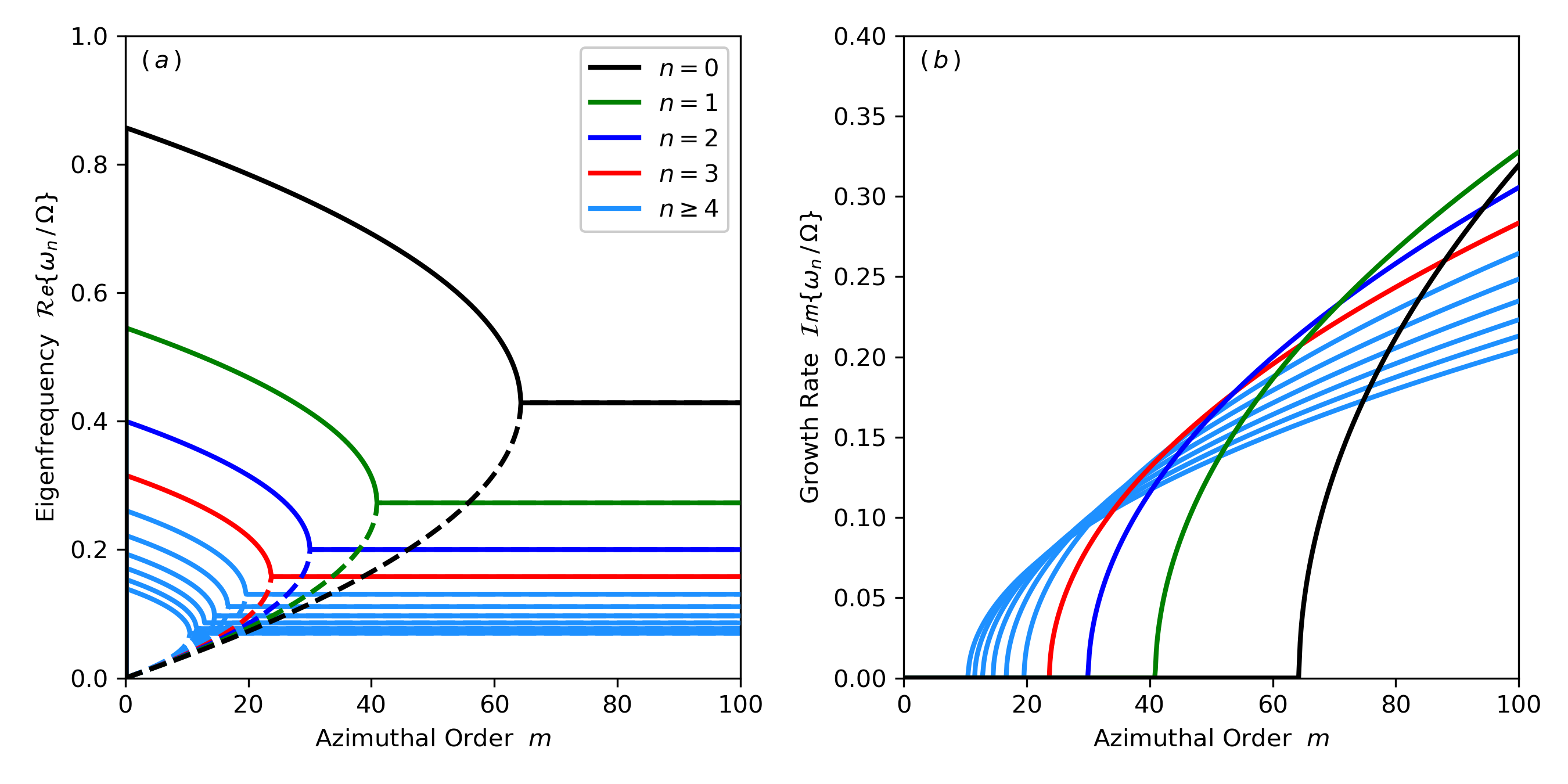}

	\caption{Complex eigenfrequencies for the semi-infinite domain. The four
modes with the lowest radial order $n$ are shown with black, green, blue, and red
curves, while all higher orders are indicated with a common light blue color. The
two panels present, as functions of the azimuthal order,
(a) the real part of the complex frequency and (b) the growth rate or the absolute
value of the imaginary part of the complex frequency. At each value of the azimuthal
order, there are two gravito-inertial wave solutions. For low azimuthal orders, both
of the gravito-inertial waves are stable and prograde propagating with purely
real frequency. The two modes have different frequencies and hence different
longitudinal phase speeds. The faster of these is shown with solid curves, while
the slow waves are drawn with dashed curves. For sufficiently large azimuthal order,
the two modes become unstable and correspond to prograde-propagating convective modes.
	\label{fig:infinite_eigenfrequencies}}
\end{figure*}

\begin{figure*}
	\epsscale{1.0}
	\plotone{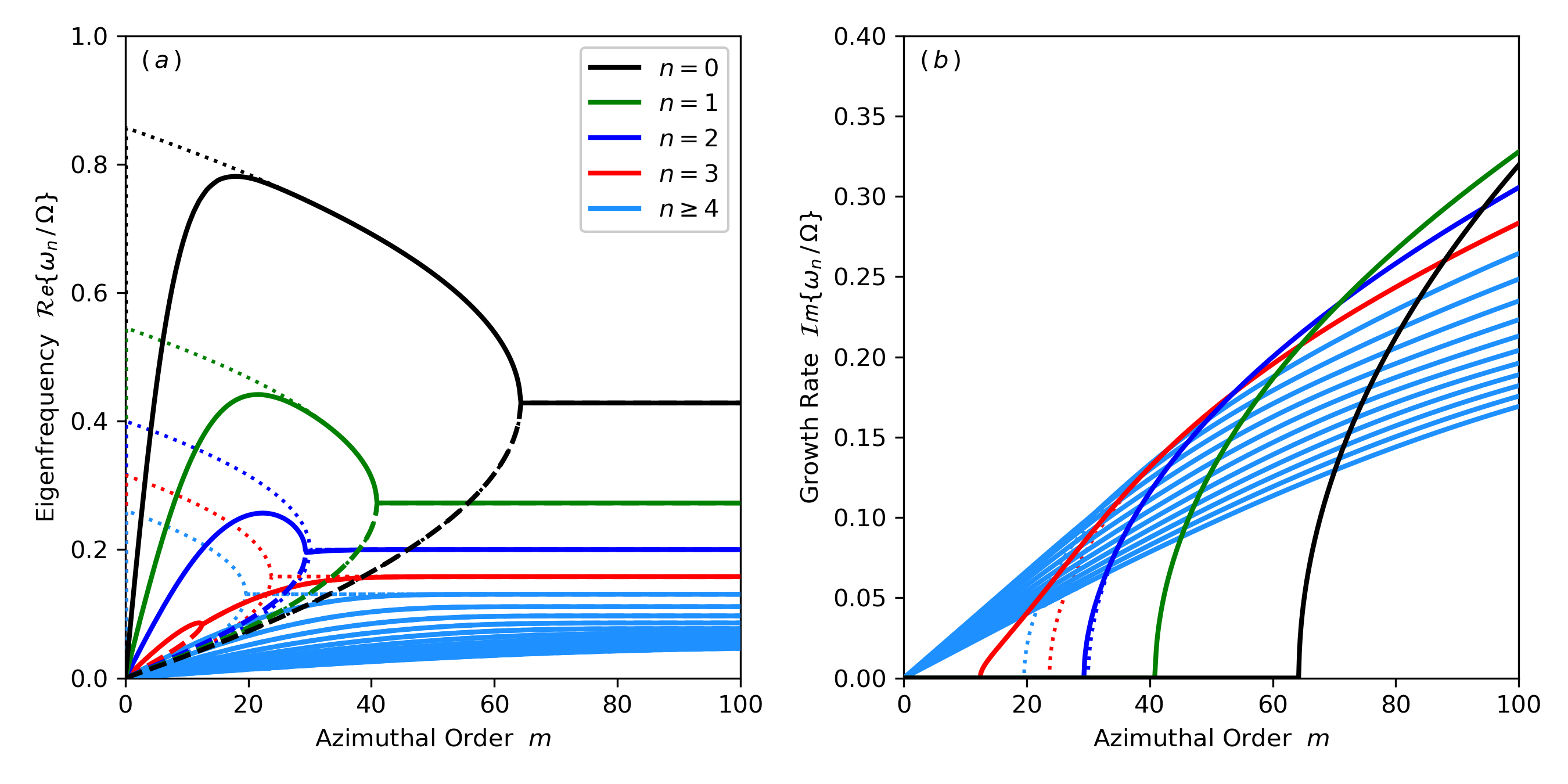}

	\caption{The complex eigenfrequencies for the finite domain are shown using
the solid and dashed curves, with the line styles and colors having the same meaning
as in Figure~\ref{fig:infinite_eigenfrequencies}. The dotted curves indicate the
same eigenfrequencies for the semi-infinite domain (see Figure~\ref{fig:infinite_eigenfrequencies}).
Only four of the lowest radial order modes can be stabilized by rotation. All higher-order
modes are unstable for all azimuthal orders. The eigenfrequencies for the two types
of domains converge for sufficiently large wavenumber.
	\label{fig:finite_eigenfrequencies}}
\end{figure*}

\noindent This behavior is exhibited in Figures~\ref{fig:infinite_eigenfrequencies}
and \ref{fig:finite_eigenfrequencies}. The polytropic atmosphere for each is weakly
unstable with $S = -5 \times 10^{-3}$ or, equivalently,
$\alpha - \hat{\alpha} = -5 \times 10^{-7}$. Figure~\ref{fig:infinite_eigenfrequencies} 
illustrates the complex eigenfrequencies for the low-frequency gravito-inertial
waves in a semi-infinite domain (see subsection \ref{subsec:semi-infinite_domain}). The
left-hand panel shows the real part of the eigenfrequencies, while the right-hand
panel presents the imaginary part or growth rate. The first four radial orders
($n=0,1,2,3$) are shown as colored curves, respectively, black, green, blue, and
red. Modes of higher radial order (up through $n=9$) are displayed in light blue.
There are two gravito-inertial wave solutions and both are prograde propagating
with positive frequencies. The higher frequency solution, or the fast gravito-inertial
wave, is shown with the solid curves.  The slow gravito-inertial wave is displayed
with dashed curves. As the wavenumber increases, the slow and fast gravito-inertial
waves have real frequencies that slowly converge. At the threshold of convergence
(at marginal stability) and beyond, the two modes become a complex conjugate pair
with the same real part of their frequencies and oppositely signed imaginary parts
(only the positive root is illustrated). When the two modes have purely real frequencies
the modes correspond to overstable convective modes. Both the fast and slow waves
would have been convectively unstable and non-oscillatory in the absence of rotation.
When the mode frequencies are complex, the two solutions correspond to unstable,
oscillatory convective modes that travel prograde at the same speed. For moderate
values of the azimuthal order $m$, these unstable modes have a growth rate that is
commensurate with their oscillation frequency. Hence, the modes grow in amplitude
over a time scale of roughly the rotation period. At marginal stability and beyond,
the two gravito-inertial waves have been called thermal Rossby waves in the convection
literature \citep[e.g.,][]{Busse:1986}. Here, we use the term thermal Rossby wave
to describe both the unstable and stable solutions.

Figure~\ref{fig:finite_eigenfrequencies} displays the complex eigenfrequencies
for the finite domain. The color of the curves and the line styles have the same
meaning  as in Figure~\ref{fig:infinite_eigenfrequencies}. We have included the
eigenfrequencies for the semi-infinite domain as the dotted curves. Only the
lowest four radial orders possess low-wavenumber overstable modes. For $n\geq4$
the gravito-inertial waves are unstable for all azimuthal orders. The lowest order
modes look very similar to those of the semi-infinite domain, except for the fast
gravito-inertial wave at the lowest wavenumbers. This occurs because the lower
turning point is below the bottom of the radial domain for low wavenumbers. Hence,
these modes are sensitive to the boundary condition at the bottom of the layer.
Similar behavior for the thermal Rossby wave where the eigenfrequency approaches
zero as the zonal wavenumber approaches zero has been seen previously
\citep{Glatzmaier:1981, Bekki:2022, Hindman:2022}. As was noted for the semi-infinite domain,
the growth rate of the unstable modes is comparable to the rotation rate of the
star for moderate azimuthal and radial orders.

%-------------------------------------------------------------------------------%
%	3.4	Eigenfunctions							%
%-------------------------------------------------------------------------------%
\subsection{Eigenfunctions}
\label{subsec:eigenfunctions}

Oddly, if the boundary conditions lack explicit dependence on the wave frequency,
$\omega$, the two acoustic modes and the two gravito-inertial wave modes have identical
eigenfunction for the Lagrangian pressure fluctuation for the same zonal wavenumber
$k_x$. This arises because the ODE for the Lagrangian pressure fluctuation depends
on the eigenfrequency only through the eigenvalue $\kappa$, and all four wave modes
have the same degenerate eigenvalue. Where the four wave modes differ is the eigenfunctions
for the other physical variables. For example, the zonal and radial velocity components, $u$ and
$w$ respectively, can be obtained from the Lagrangian pressure fluctuation through
the following differential operators \citep{Hindman:2022},

\begin{eqnarray}
	\label{eqn:u_operator}
	u &=& \frac{\sigma^2 \omega}{\omega^4 - \sigma^4} \left[\frac{d}{dz} + \frac{\omega^2 k_x}{\sigma^2} - \frac{1}{H}\right] \left(\frac{\delta P}{\rho_0}\right)\; ,
\\
	\label{eqn:w_operator}
	w &=& -\frac{i\omega^3}{\omega^4 - \sigma^4} \left[\frac{d}{dz} + \frac{\sigma^2 k_x}{\omega^2} - \frac{1}{H}\right] \left(\frac{\delta P}{\rho_0}\right)\; ,
\end{eqnarray}

\noindent where $\sigma^2 \equiv gk_x -2\Omega\omega$. Since, these differential
operators are frequency dependent, the velocity eigenfunctions for the acoustic
waves and the gravito-inertial waves of the same radial order $n$ will differ even
though they possess the same eigenvalue $\kappa_n$.

Further, in the low-frequency limit that is appropriate for the two gravito-inertial
modes, the radial velocity, $w$, is proportional to the reduced Lagrangian Pressure
fluctuation $w \propto \delta P / \rho_0$ and the zonal velocity, $u$, is identical
to within a constant amplitude for the fast and slow gravito-inertial waves. This
becomes apparent when the low-frequency limit, $\epsilon \ll 1$, of
Equations~\eqnref{eqn:u_operator} and \eqnref{eqn:w_operator} is taken,

\begin{eqnarray}
	\label{eqn:u_lowfreq}
	u &\to& -\frac{\omega}{gk_x} \left[\frac{d}{dz} - \frac{1}{H}\right] \left(\frac{\delta P}{\rho_0}\right)\; ,
\\
	\label{eqn:w_lowfreq}
	w &\to& \frac{i\omega}{g} \left(\frac{\delta P}{\rho_0}\right)\; .
\end{eqnarray}

Finally, we note that even for the unstable modes the Lagrangian pressure fluctuation
has a real eigenfunction and from Equations~\eqnref{eqn:u_lowfreq} and \eqnref{eqn:w_lowfreq}
we can see that the eigenfunctions for the two velocity components are complex, but
possess a complex phase that is independent of height. Furthermore, since we have
adopted a weakly unstable polytrope with $S=-5 \times 10^{-3}$ and $\alpha - \hat{\alpha} = -5 \times 10^{-7}$,
the eigenfunctions are nearly indistinguishable from the eigenfunctions for a neutrally
stable atmosphere ($S = 0$ and $\alpha = \hat{\alpha}$). Hence, instead of providing
redundant illustrations here, we refer the reader to Figures 5 and 7 of \cite{Hindman:2022}.
Figure 5 from \cite{Hindman:2022} shows eigenfunctions for the semi-infinite atmosphere
developed in ~\ref{subsec:semi-infinite_domain} and Figure 7 presents eigenfunctions
for the finite layer discussed in subsection~\ref{subsec:finite_domain}

%%%%%%%%%%%%%%%%%%%%%%%%%%%%%%%%%%%%%%%%%%%%%%%%%%%%%%%%%%%%%%%%%%%%%%%%%%%%%%%%%
%										%
%       4.	Coupling of Thermal Rossby Waves with Prograde g Modes		%
%										%
%%%%%%%%%%%%%%%%%%%%%%%%%%%%%%%%%%%%%%%%%%%%%%%%%%%%%%%%%%%%%%%%%%%%%%%%%%%%%%%%%
\section{Coupling of Thermal Rossby Waves with Prograde g Modes}
\label{sec:Mode_Coupling}

In the previous section we applied a perfectly-reflecting boundary condition (i.e.,
$\delta P = 0$) at the bottom of the convection zone. Hence, the thermal Rossby modes
were completely confined to the convection zone and any potential coupling to the
$g$ modes of the star's stably-stratified radiative interior was neglected. Since,
the square of the buoyancy frequency increases dramatically over the thin boundary
between the convection zone and radiative interior, we expect that the reflection
is indeed almost total and the coupling between $g$ modes of the interior and
thermal Rossby modes of the convection zone is extremely weak, except when the
thermal Rossby wave and the gravity wave happen to have a common frequency. Hence,
in a dispersion diagram we should see two distinct families of dispersion curves and
where those curves cross (i.e., have common frequencies) we expect to see avoided
crossings.

In order to demonstrate how the avoided crossings might appear, we have developed
a simple, illustrative model consisting of a weakly-unstable polytropic layer of
finite depth $D$ that overlays an isothermal layer of thickness $L$. The polytrope
represents the Sun's convection zone and the isothermal atmosphere its radiative
interior. As before, we adopt $D = 200$ Mm and $S = -5 \times 10^{-3}$. We choose
the buoyancy frequency of the isothermal atmosphere such that it is comparable to
the buoyancy frequency of the Sun's interior, $N = 10^{-3}$ s$^{-1}$. For illustrative
purposes we set the depth of the isothermal layer $L$ to an unrealistically thin
value, $L = 500$ km. We do this to control the frequency spacing between $g$ modes
of nearby radial order. If we were to use a more realistic value, say $L = 500$ Mm,
the separation between $g$ modes would be so tiny that the density of $g$-mode ridges
in the dispersion diagram would obscure the effect that we are seeking to illustrate.
The reader should note that the fine spacing between $g$ modes is not the result
of our choice of an isothermal atmosphere.  Instead, it is a consequence of the smallness
of inertial wave frequencies in comparison to the radiative interior's buoyancy frequency,
$\omega \ll N$.

For boundary conditions, we require that the  solution is regular at the origin, $z=0$,
and that the vertical derivative of the Lagrangian pressure fluctuation vanishes at
the bottom of the isothermal atmosphere, $z = - (D+L)$. This later condition is appropriate
if the bottom boundary were to correspond to the radial center of the star. At the interface
between the polytrope and isothermal atmosphere ($z = -D$), we require that the Lagrangian
pressure fluctuation and its radial derivative are continuous. These conditions are
consistent with the continuity of the  pressure fluctuation and the normal velocity
component. In the isothermal atmosphere the solutions for $\Psi$ are sinusoidal
\citep[see][]{Hindman:2022} and the boundary condition at the bottom of that layer
fixes the phase of the sinusoid. Within the polytrope, the upper boundary condition
fixes the solution to be the one proportional to the $M$ Kummer function. The
eigenfunction can therefore be written in the form

\begin{equation}
	\Psi(z) = \left\{
	\begin{array}{ll}
		C_{\rm iso} \, \cos\left[K_{\rm iso}(z+D+L)\right] & {\rm for~} z<-D  \; ,
\\
		C_{\rm poly} \, z^\nu \, e^{k_x z} \, M\left(\nu - \kappa, 2\nu, -2k_x z\right) & {\rm for~} z>-D  \; ,
	\end{array}
	\right.
\end{equation}

\noindent where $C_{\rm iso}$ and $C_{\rm poly}$ are constant amplitudes and $K_{\rm iso}$
is the radial wavenumber within the isothermal atmosphere,

\begin{equation}
	K_{\rm iso}^2 = \frac{\omega^2 - \omega_c^2 - 4 \Omega^2}{c^2} - k_x^2 \left(1 - \frac{N^2}{\omega^2}\right)
			+ \frac{2\Omega k_x}{\omega {\cal H}} \; .
\end{equation}

\noindent The global dispersion relation arises from the two continuity conditions and has the form,

\begin{equation}
	K_{\rm iso} \tan\left(K_{\rm iso} L\right) + k_x - \frac{\alpha+2}{2D} - 2k_x \frac{M'\left(\nu - \kappa_n, 2\nu, 2 k_x D\right)}{M\left(\nu - \kappa_n, 2\nu, 2 k_x D\right)}=0 \; ,
\end{equation}

\noindent where $M'=dM/d\zeta$ is the derivative of the $M$ Kummer function with
respect to its third argument (the spatially varying argument).

\begin{figure*}
	\epsscale{1.0}
	\plotone{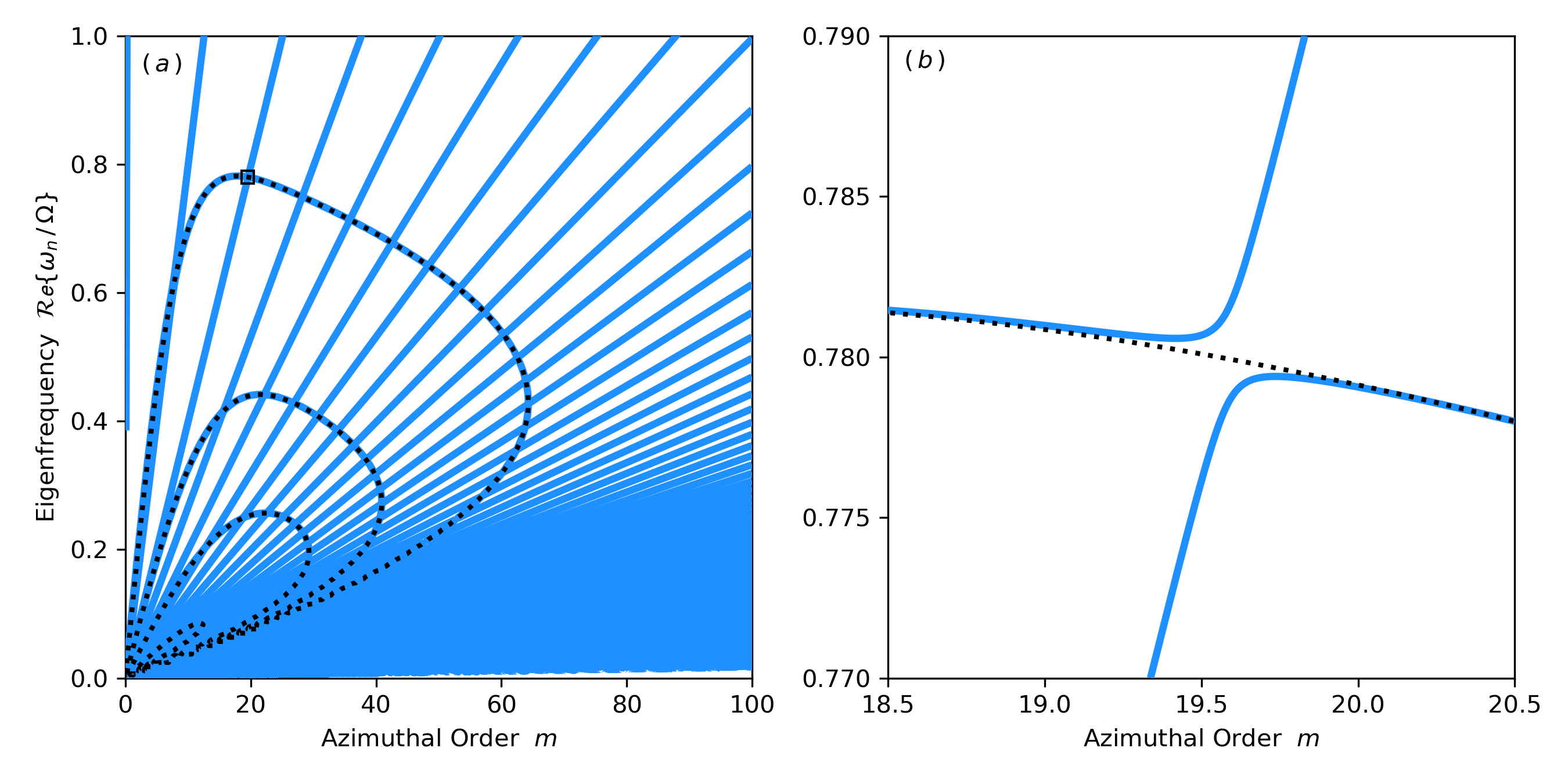}

	\caption{The real part of the complex eigenfrequencies for the two-layer model,
consisting of a weakly unstable polytropic convection zone that overlies a stably
stratified isothermal radiative interior. (a) Two families of solutions are possible
(shown with blue curves, the gravito-inertial waves that reside in the convection
zone and the $g$ modes that are confined to the radiative zone. The black dotted
lines indicate four of the gravito-inertial waves that we obtained previously when
a perfectly reflecting boundary was placed at the bottom of the convection zone. These
are the same eigenfrequencies that were illustrated in Figure~\ref{fig:finite_eigenfrequencies}.
In the two-layer model, the interface between the convection zone and the radiative
interior is partially transmitting and the two wave cavities can communicate. Because
of this, when a gravito-inertial wave has the same frequency as a $g$ mode there is
an avoided crossing. (b) A zoom-in view of the avoided crossing marked by the small
black box in panel a.
	\label{fig:avoided_crossings}}
\end{figure*}

Figure~\ref{fig:avoided_crossings}$a$ illustrates the eigenfrequencies for this
``toy" model. As predicted there are two families of solutions. Superimposed over
the family of thermal Rossby waves that we found in the previous section for the
finite domain (see Figure~\ref{fig:finite_eigenfrequencies}), one observes a set
of $g$ modes that have frequencies that monotonically increase with the zonal
wavenumber $k_x$. The thermal Rossby waves that we found before by ignoring the
presence of the $g$ mode cavity are indicated by the black dotted curves. Wherever
the dispersion curve for a thermal Rossby mode trapped in the convection zone
crosses a dispersion curve for an interior $g$ mode an avoided crossing occurs.
Since, the jump in the buoyancy frequency between the isothermal atmosphere and
the polytrope is strong with a ratio of $7\times 10^8$, the reflection coefficient
between the two layers is nearly unity and the avoided crossings are extremely tight.
Therefore, they are not easily visible in panel $a$.
Figure~\ref{fig:avoided_crossings}$b$ provides a zoom-in view of the crossing
indicated in panel $a$ by the small square box. As can be seen, the coupling between
a $g$ mode and a thermal Rossby waves occurs over a very narrow range of frequencies
where the two distinct cavities have a common resonance.

%%%%%%%%%%%%%%%%%%%%%%%%%%%%%%%%%%%%%%%%%%%%%%%%%%%%%%%%%%%%%%%%%%%%%%%%%%%%%%%%%
%										%
%       5.      Discussion		   					%
%										%
%%%%%%%%%%%%%%%%%%%%%%%%%%%%%%%%%%%%%%%%%%%%%%%%%%%%%%%%%%%%%%%%%%%%%%%%%%%%%%%%%

\section{Discussion}
\label{sec:Discussion}

We have developed an analytic solution for atmospheric waves of all types that is
valid in a compressible, polytropically-stratified atmosphere. In total there are
four atmospheric wave solutions: two high-frequency acoustic waves and two low-frequency
gravito-inertial waves. Our analytic solution has the form of a longitudinal plane
wave multiplied by a radial eigenfunction comprised of Kummer functions. This
solution is valid for any value of the polytropic index, independent of whether the
polytrope is stable or unstable to convective overturning. We have explicitly illustrated
solutions only for a weakly unstable stratification appropriate for a stellar convection
zone. In such a stratification, both gravito-inertial waves propagate in the prograde
direction. In a neutrally stable atmosphere ($N^2 \to 0$), the slow wave becomes a
degenerate, zero-frequency wave that is stationary in the rotating frame and the fast
wave persists as a prograde-propagating wave \citep{Hindman:2022}.

Previous studies have explicitly considered low-frequency waves in a neutrally
stable atmosphere. \cite{Glatzmaier:1981} solved for the eigenfrequencies in the
anelastic limit for a polytropic layer of finite radial extent using Frobenius
expansions to describe the eigenfunctions. \cite{Bekki:2022} solved for the linear
eigenmodes by numerically solving the spatially discretized fluid equations in
spherical geometry. \cite{Hindman:2022} derived the Kummer-function solutions
in the limit of neutral stability ($N^2 = 0$). For all of these previous efforts,
the wave solutions correspond to the fast gravito-inertial wave that we have explored
here, which becomes a pure inertial wave when the buoyancy frequency is identically
zero throughout the radial domain. 

%-------------------------------------------------------------------------------%
%	5.1	Wave Nomenclature						%
%-------------------------------------------------------------------------------%
\subsection{Wave Nomenclature}
\label{subsec:wave_nomenclature}

The two gravito-inertial waves that we find here, and which we have called the fast
and slow thermal Rossby wave, are completely analogous to the two wave solutions
found by \cite{Busse:1986} for a Boussinesq fluid in a rotating cylindrical shell.
Just as we have assumed here, \cite{Busse:1986} ignored variations and motions parallel
to the rotation axis. While we have a compressional $\beta$ effect due to the gravitational
stratification of density, \cite{Busse:1986} introduced a topological $\beta$-effect
by allowing the upper and lower caps of the cylindrical domain to be conical with
a tilt designed to mimic the effects of a spherical surface. To illustrate the similarities,
we explore Busse's model in the dissipationless limit.

\cite{Busse:1986} derived dispersion relations for linear waves for systems with
and without diffusion. For an inviscid fluid without thermal conduction, the two
wave solutions found by \cite{Busse:1986} possess frequencies that can be rewritten
in the following form

\begin{equation} \label{eqn:Busse}
	\omega = \frac{K k_x \tau^{-1}}{k_x^2 +k_z^2} 
		\left[	\left(1+\frac{k_z^2}{K^2}\right)^{1/2} \pm
			\left(1-\frac{k_x^2}{K^2}\right)^{1/2}\right] \; ,
\end{equation}

\noindent where $k_x$ is the azimuthal wavenumber, $k_z$ is the radial wavenumber,
$\tau$ is a free-fall time, and $K$ is the azimuthal wavenumber that corresponds to
the margin of stability,

\begin{eqnarray}
	\tau &\equiv& \frac{D}{g \alpha_T \Delta T}^{1/2} \; ,
\\
	K^2  &\equiv& 4\sin^2(\chi) \frac{\tau^2\Omega^2}{D^2} - k_z^2 \; .
\end{eqnarray}

\noindent The free-fall time $\tau$ and the threshold wavenumber $K$, depend on
the thickness of the cylindrical shell $D$, the coefficient of thermal expansion
$\alpha_T$, the temperature difference between the inner and outer cylindrical
surfaces $\Delta T$, and the angle $\chi$ between the upper and lower conical
surfaces and the equatorial plane. Note, we have written the centrifugal buoyancy
term that appears in Busse's expressions in the form of a more traditional gravitational
buoyancy.

\begin{figure*}
	\epsscale{1.0}
	\plotone{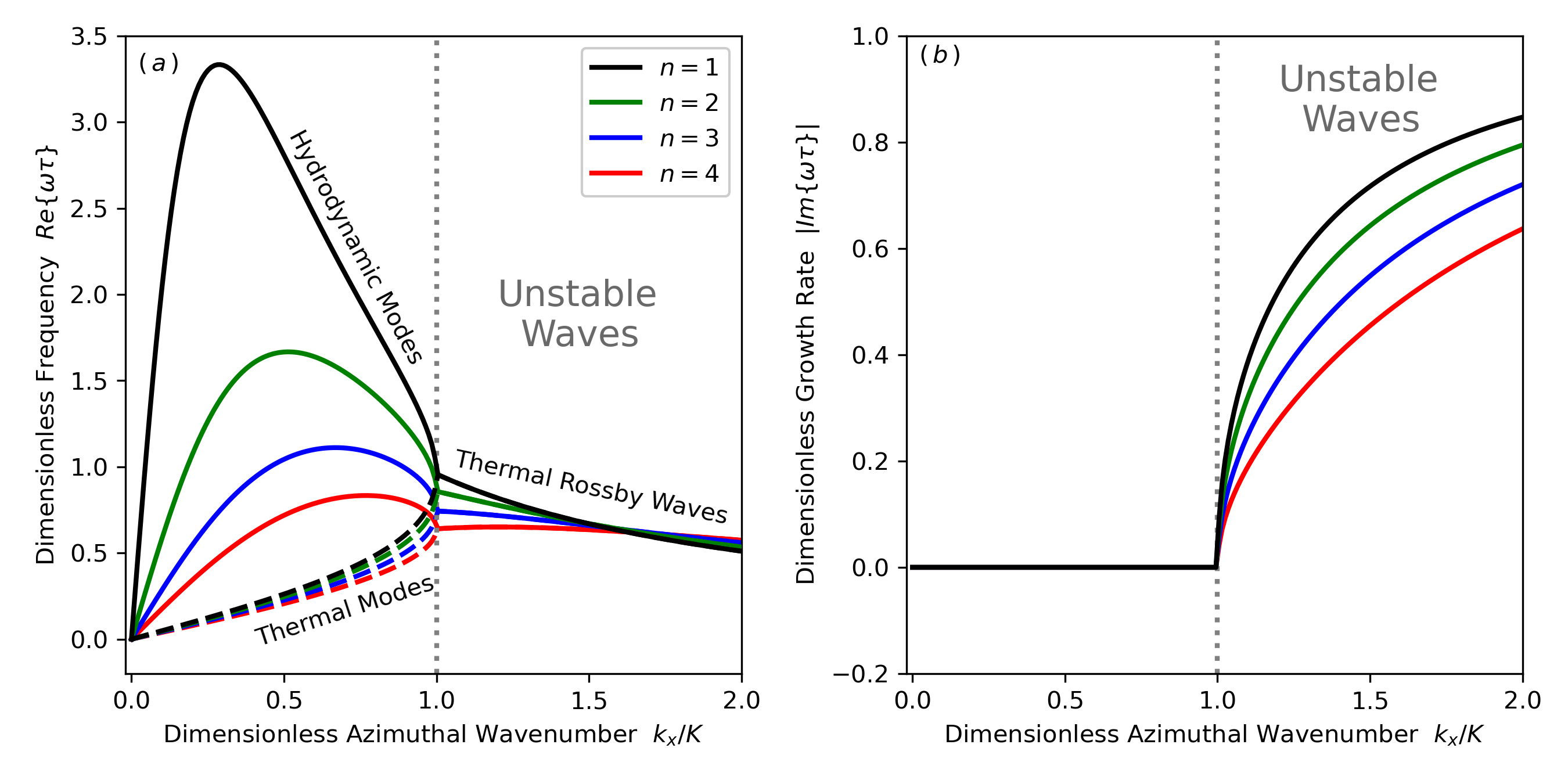}

	\caption{The complex eigenfrequencies for the cylindrical, Boussinesq model
of \cite{Busse:1986}. The solid and dashed curves indicate the fast and slow gravito-inertial
waves, respectively. \cite{Busse:1986} referred to the fast waves as the hydrodynamic
modes and the slow waves as the thermal modes. The unstable modes (and marginally stable
modes) were collectively called thermal Rossby waves. The colors correspond to radial
order $n$ of the mode as indicated in the legend. Only the four lowest radial orders
are plotted. These modes are completely analogous to the gravito-inertial waves that
we have explored here (see Figure~\ref{fig:finite_eigenfrequencies}).  The $\beta$
effect is topological in Busse's model and compressional in ours.  Both lead to prograde
propagating waves.
	\label{fig:Busse}}
\end{figure*}

In Figure~\ref{fig:Busse} we plot the real and imaginary parts of the eigenfrequencies
for these two modes for the parameter value $KD/\pi = 0.3$. All of these solutions
have a purely real frequency for $k_x \leq K$ and a complex frequency for $k_x > K$.
The vertical gray line indicates the marginal wavenumber $k_x=K$ that separates the
stable and unstable solutions. The color of each curve indicates the radial order $n$
of the mode with $k_z = n\pi/D$. Note, the similarity between Figures~\ref{fig:finite_eigenfrequencies}
and \ref{fig:Busse}. Each model has two gravito-inertial waves, and both modes are 
stable for low azimuthal wavenumber and become unstable once the azimuthal wavenumber
crosses a threshold value. The similarities arise because the waves in both models
are controlled by a $\beta$ effect, the effect being topological in Busse's and
compressional in ours. 

The solid curves in Figure~\ref{fig:Busse} indicate the wave with the faster phase
speed, i.e., the plus sign in Equation~\eqnref{eqn:Busse}. When the mode is stable,
\cite{Busse:1986} refers to this mode as the ``hydrodynamic mode" and notes that it
behaves like a Rossby wave. In the limit of rapid rotation or weak instability,
$\Omega \tau \to \infty$, this fast wave satisfies the dispersion relation

\begin{equation}
	\omega = \frac{\beta k_x}{k_x^2 + k_z^2} + \cdots \; ,
\end{equation}

\noindent where the topological $\beta$ effect is given by $\beta = 4\sin(\chi)\Omega/D$.
In our model, the same limit of rapid rotation or weak instability is achieved by
considering $N^2 / \Omega^2 \to 0$. In \cite{Hindman:2022} we demonstrate that
the local dispersion relation for the fast mode reduces to

\begin{equation}
	\omega = \frac{\beta k_x}{k_x^2 + k_z^2 + k_c^2} + \cdots \; ,
\end{equation}

\noindent with a compressional $\beta$ effect given by $\beta = 2 \Omega/{\cal H}$.
This is clearly a Rossby wave with a correction for stratification (appearing in
$\beta$ and through $k_c^2$).

The slower of the two modes---i.e., the minus sign in Equation~\eqnref{eqn:Busse}---is
illustrated in Figure~\ref{fig:Busse} with dashed curves. \cite{Busse:1986} calls the
stable slow mode the ``thermal mode" and points out that it becomes a zero-frequency
wave in the limit of rapid rotation (or weak instability),

\begin{equation}
	\omega  = \frac{\csc(\chi) D}{2 \Omega^{1/2} \tau^{3/2}} k_x + \cdots \; .
\end{equation}

\noindent This same behavior occurs for our slow gravito-inertial wave. In the rapidly
rotating limit, \cite{Hindman:2022} derived the following local dispersion relation
for the slow mode,

\begin{equation}
	\omega = \frac{{\cal H} |N^2|}{\Omega} k_x + \cdots \; .
\end{equation}

Finally, to complete the discussion of Busse's terminology, he identifies both
wave branches as thermal Rossby waves, but does so only during a discussion of the
marginally-stable and unstable modes. We adopt his nomenclature and refer to both
gravito-inertial waves as thermal Rossby waves, but for ease of language, we have
chosen to use the term thermal Rossby wave, irrespective of the stability of the
mode.

%-------------------------------------------------------------------------------%
%	5.2	The Stability Threshold						%
%-------------------------------------------------------------------------------%
\subsection{The Stability Threshold}
\label{subsec:stability_threshold}

We have calculated the eigenmodes for a weakly unstable polytrope with the explicit
intention of exploring the behavior of overstable convective modes in a star's
convection zone. In the absence of rotation, the internal gravity waves of such an
atmosphere would take on the form of unstable convective modes. We confirm
the well-known property that rotation can stabilize convective motions, resulting in
overstable convective modes that propagate in the prograde direction. Only the longest 
wavelengths are stabilized and any waves with an azimuthal order larger than

\begin{equation}
	m > m_{\rm thresh} = -\frac{\alpha^2 \epsilon}{4 \kappa_n S} \; ,
\end{equation}

\noindent are unstable, taking on the form of prograde-propagating, oscillating,
convective modes. This threshold wavenumber corresponds to the location in the dispersion
diagram where the frequencies of the fast and slow gravito-inertial waves merge and
become complex conjugates (see Figures~\ref{fig:infinite_eigenfrequencies} and
\ref{fig:finite_eigenfrequencies}).

One can discern that this wavenumber threshold is highly sensitive to the superadiabatic
gradient of the convection zone. This is most easily recognized by expressing the threshold
in terms of the atmosphere's density scale height $H$, its specific entropy gradient,
$ds/dr$, and the specific heat at constant pressure $c_p$,

\begin{equation}
	m > -\frac{\epsilon^2 c_p}{2(\gamma-1)\kappa_n} \left(H \frac{ds}{dr}\right)^{-1} \; .
\end{equation}

\noindent In deriving this expression, we have assumed that the atmosphere is nearly
adiabatic, $\alpha \approx \left(\gamma-1\right)^{-1}$, utilized the relationship
between the entropy gradient and the buoyancy frequency

\begin{equation}
	\frac{ds}{dr} = \frac{c_p N^2}{g} \; ,
\end{equation}

\noindent and used \eqnref{eqn:S_def} to express the buoyancy frequency in terms
of our stability parameter $S$,

\begin{equation}
	N^2 = \frac{2 \epsilon}{\alpha} \frac{gS}{H} \; .
\end{equation}

If the frequencies of the overstable convective modes were to be observable, and the
wavenumber threshold $m_{\rm thresh}$ measurable, the threshold could be used to estimate
the superadiabaticity of the convection zone,

\begin{equation}
	\frac{ds}{dr} \approx -\frac{\epsilon^2 c_p}{2(\gamma-1)\kappa_n m_{\rm thresh}} \frac{1}{H} \; .
\end{equation}

%-------------------------------------------------------------------------------%
%	5.3	Gravity Mode Visibility						%
%-------------------------------------------------------------------------------%
\subsection{Gravity Mode Visibility}
\label{subsec:mode_visibility}

The $p$ modes and $f$ modes of helioseismology have been observed and utilized
since the 1960s to probe the properties of the solar interior. These modes have a
high visibility because their wave cavities reside in the convection zone and their
eigenfunctions extend very close to the solar photosphere where motions and intensity
fluctuations associated with the modes can be directly observed. On the other hand
the $g$ modes that live within the radiative interior have not been unambiguously
detected. Several claims of such detection have been made over the years
\citep[e.g.,][]{Gabriel:2002, Turck-Chieze:2004, Garcia:2007, Fossat:2017}; however,
the measurements are difficult and independent verification has proven slippery. The
primary reason for the elusivity of the $g$ modes is their low amplitude in the photosphere.
Their cavity is confined to the radiative interior and the eigenfunctions undergo many
decay lengths between the upper boundary of their cavity and the photosphere.

Our findings here suggest that the prograde $g$ modes might be more visible than
their  retrograde cousins. These prograde modes have the possibility to couple with
thermal Rossby waves which have a shallower wave cavity. This cavity should occupy
the lower half of the convection zone and therefore the distance between the upper
turning point of the mode and the height of observation in the photosphere is roughly
half of the distance for the retrograde modes. This property is analogous to the
mixed modes of a red giant for which a $g$ mode of the interior couples to a $p$
mode of the envelope. Those frequencies where the thermal Rossby wave and $g$ modes
have a common frequency might allow the properties of the $g$ mode cavity to be
explored. Of course, thermal Rossby waves would need to be detected first. But the
observational discovery of the Sun's resonant inertial waves is still relatively
recent \citep[e.g.,][]{Loeptien:2018, Hanasoge:2019, Gizon:2021} and many additional
discoveries are likely to occur.

%%%%%%%%%%%%%%%%%%%%%%%%%%%%%%%%%%%%%%%%%%%%%%%%%%%%%%%%%%%%%%%%%%%%%%%%%%%%%%%%%
%                                                                               %
%       Acknowledgments                                                         %
%                                                                               %
%%%%%%%%%%%%%%%%%%%%%%%%%%%%%%%%%%%%%%%%%%%%%%%%%%%%%%%%%%%%%%%%%%%%%%%%%%%%%%%%%

\begin{acknowledgments}
    This work was supported by NASA through grants 80NSSC17K0008, 80NSSC18K1125, 80NSSC19K0267, and 80NSSC20K0193. R.J. would like to acknowledge the support of MSRC (SoMaS), University of Sheffield (UK). B.W.H. thanks Jonathon Aurnou, Nicholas Featherstone, and Keith Julien for useful discussions that helped to shape the work presented here. This work was done in collaboration with the COFFIES DSC.
\end{acknowledgments}

%%%%%%%%%%%%%%%%%%%%%%%%%%%%%%%%%%%%%%%%%%%%%%%%%%%%%%%%%%%%%%%%%%%%%%%%%%%%%%%%%
%										%
%       References						   		%
%										%
%%%%%%%%%%%%%%%%%%%%%%%%%%%%%%%%%%%%%%%%%%%%%%%%%%%%%%%%%%%%%%%%%%%%%%%%%%%%%%%%%

%%%%%%%%%%%%%%%%%%%%%%%%%%%%%%%%%%%%%%%%%%%%%%%%%%%%%%%%%%%%%%%%%%%%%%%%%%%%%%%%%
%										%
%       Figures  				                    		%
%										%
%%%%%%%%%%%%%%%%%%%%%%%%%%%%%%%%%%%%%%%%%%%%%%%%%%%%%%%%%%%%%%%%%%%%%%%%%%%%%%%%%

%%%%%%%%%%%%%%%%%%%%%%%%%%%%%%%%%%%%%%%%%%%%%%%%%%%%%%%%%%%%%%%%%%%%%%%%%%%%%%%%%
%		Figure 1 - Eigenvalue kappa

%%%%%%%%%%%%%%%%%%%%%%%%%%%%%%%%%%%%%%%%%%%%%%%%%%%%%%%%%%%%%%%%%%%%%%%%%%%%%%%%%
%		Figure 2 - Eigenvalue kappa

%%%%%%%%%%%%%%%%%%%%%%%%%%%%%%%%%%%%%%%%%%%%%%%%%%%%%%%%%%%%%%%%%%%%%%%%%%%%%%%%%
%		Figure 3 - Eigenfrequencies - Infinite Domain

%%%%%%%%%%%%%%%%%%%%%%%%%%%%%%%%%%%%%%%%%%%%%%%%%%%%%%%%%%%%%%%%%%%%%%%%%%%%%%%%%
%		Figure 4 - Eigenfrequencies - Finite Domain

%%%%%%%%%%%%%%%%%%%%%%%%%%%%%%%%%%%%%%%%%%%%%%%%%%%%%%%%%%%%%%%%%%%%%%%%%%%%%%%%%
%		Figure 5 - Avoided Crossings

%%%%%%%%%%%%%%%%%%%%%%%%%%%%%%%%%%%%%%%%%%%%%%%%%%%%%%%%%%%%%%%%%%%%%%%%%%%%%%%%%
%		Figure 6 - Eigenfrequencies - Busse's Model

\end{document}